\documentclass{aastex}
\shortauthors{Dinh-V-Trung et al.}
\shorttitle{Nature of ULIRG...}
\usepackage{epsfig,psfig}

\begin{document}

\title{Nature of widely separated ultraluminous infrared galaxies}

\author{Dinh-V-Trung\altaffilmark{1}, K.Y. Lo, D.-C. Kim}
\affil{Institute of Astronomy and Astrophysics, Academia Sinica, P.O. Box 1-87,
Nankang, Taipei 11529, Taiwan, ROC.}
\email{trung@asiaa.sinica.edu.tw, kyl@asiaa.sinica.edu.tw, kim@asiaa.sinica.edu.tw}
\author{Yu Gao}
\affil{Infrared Processing and Analysis Center, MS 100-22, California Institute of Technology, Pasadena, CA 91125.}
\email{gao@ipac.caltech.edu}
\and
\author{R.A. Gruendl}
\affil{Department of Astronomy, University of Illinois at Urbana-Champaign, Urbana, IL 61801.}
\email{gruendl@astro.uiuc.edu}
\altaffiltext{1}{On leave from Institute of Physics, P.O. Box 429 BoHo 10000, Hanoi, VietNam}
\begin{abstract}
In the complete sample of ultraluminous infrared galaxies (ULIRGs) compiled by Kim (1995) 
about 5\% consists of widely separated galaxies which are presumably in the early phase of interaction. 
This fact is contrary to the conventional view that ULIRGs are in the 
final stages of the merger of two gas-rich disk galaxies. 
We have undertaken high resolution CO(J=1-0) observations for the
ultraluminous infrared galaxies that have nuclear separations larger than 20 kpc.
We have detected the CO emission in 5 out of 6 systems, but only in 
one component of the ULIRG pairs. 4 of them have LINER spectral type and 1 galaxy has 
Seyfert II spectral type. In K'-band images these components are also brighter
than the other components which have either HII-region spectra or no detectable emission lines. 
Using the standard conversion factor, the molecular gas content is estimated to 
be a few times 10 $^{10}$ M$_\odot$, similar to that of the other ultraluminous galaxies. The result indicates that
the galaxy containing the molecular gas is also the source of most, if not all, of the 
huge far-infrared luminosity of the
system. The optical and K'-band imaging observations and optical spectra
suggest multiple merger scenario for 1 system.
If the remaining systems are in an early stage of a binary tidal interaction,
the commonly accepted interpretation of the ULIRG phenomenon as the final merger stage of two disk
galaxies may need to be re-examined.
\end{abstract}

\keywords{galaxies: interactions --- galaxies: ISM --- galaxies: starburst --- ISM: molecules}

\section{Introduction}
Ultraluminous infrared galaxies (ULIRGs) are known to emit the bulk of their energy in the infrared and
their infrared luminosities ($L_{IR} \ge 10^{12}L_{\odot}$) are equivalent to the bolometric luminosities
of optically selected quasars.
For the past decade, there have been extensive studies of the nature of these ULIRGs (Sanders \& Mirabel 1996).
Imaging studies at optical and near-infrared wavelengths show that most of these galaxies are
either interacting or merging (Sanders et al. 1988, Zou et al. 1991, Leech et al. 1994, Murphy et al. 1996,
Borne et al. 2000, Kim et al. 2001). The fraction of ULIRGs in interacting or merging ranges from at least 61\% 
(25 objects out of 41, Zou et al. 1991) to possibly as high as 100\% 
for all 10 nearby ULIRGs (Sanders et al. 1988). The rather small fraction of interacting systems 
derived by Zou et al. (1991) resulted from the use of POSS images
in which faint tidal tails are almost impossible to detect. Recent imaging data by Kim et al. (2001) showed that
close to 100\% of ULIRGs in the complete 1 Jy sample have faint tidal tails, a signature of recent interaction and merger.

One of the most basic issues of ULIRGs is the energy source powering the enormous amount of their infrared radiation.
Spectroscopic studies in the optical, near-infrared, and mid-infrared suggest that about 70\% and 30\% 
of these galaxies are powered by starburst activities and active galactic nuclei (AGNs)
respectively (Veilleux et al. 1999a, Veilleux et al. 1999b, Genzel et al. 1998a).
The huge infrared luminosity of the ULIRGs suggests that they contain a large reservoir of molecular
gas to provide fuel for the star formation activity and/or a central AGN.
Extensive searches for CO emission have indeed shown that all ULIRGs possess a huge amount of
molecular gas ($\sim$ 10$^{10}$ M$_\odot$) and that there is a trend of increasing molecular gas content with
the far-infrared emission (Downes et al. 1993, Solomon et al. 1997). 
Recent interferometric observations of nearby ULIRGs (Downes \& Solomon 1998) 
and luminous infrared galaxies (Bryant \& Scoville 1999) 
have determined that the molecular gas is concentrated in a 
circumnuclear disk of typically 1 kpc in size.
From the interferometric observations of double-nucleus ULIRGs with projected nuclear separations of 3-5 kpc,
Evans et al. (1999, 2000) find that one of the interacting pair contains the majority of molecular gas in warm ULIRGs
whereas both components contain molecular gas in cool ULIRGs.
Based on observational evidence, Sanders et al. (1988) proposed that ULIRGs originate from the merger of two
gas-rich spiral galaxies. The large amount of molecular gas accumulated in the center of the galaxies due to
the gravitational interaction can sustain strong star formation activities as well as provide
fuel to the central AGN. As the merger proceeds, the molecular gas reservoir is eventually depleted
by the star formation (Gao \& Solomon 1999) and dispersed by the AGN activities and supernovae. Finally the AGN reveals itself as an
optical quasar. The estimated lifetime of the ULIRG phase is 
about 10$^7$ to 10$^{8}$ yr. Genzel et al. (1998b) reviewed most recent observational evidence in favor of
this scenario.

Numerical simulations of interacting galaxies have shown that violent star formation activity can be triggered in
the merger of two massive disk galaxies (Barnes \& Hernquist 1996, Mihos \& Hernquist 1996, Springel 2000). Furthermore, star formation 
activities are found to be dependent mainly on the galaxy structure and the orbital geometry plays only a 
modest role (Mihos \& Hernquist 1994, 1996).
They suggest that galaxies with dense central bulges show strong star formation activities at the final stages
of merging, while bulgeless galaxies show weaker star formation activity at the early stage of interacting
and do not show any subsequent strong star formation activities. Thus, according to the numerical simulations,
the majority of ULIRGs would be in the final stages of merger.

Recently, Kim et al. (2001) have completed an optical and near-infrared imaging survey of the 
IRAS 1 Jy sample of ULIRGs (Kim \& Sanders 1998) and find that all of the ULIRGs show signs of tidal interaction
and therefore are in a merging or interacting process (Kim et al. 2001).
More interestingly, we also find that 6 out of 118 ULIRGs have projected nuclear separation larger than 20 kpc.
The large nuclear separation observed in these ULIRGs seems to contradict the conventional view of the origin of ULIRGs
and the numerical simulation results outlined above. To become an ULIRG in the early phase of interaction,
the galaxies must have accumulated a large amount of molecular gas and it is still unclear how the star formation activities
are triggered in these cases.\\
The position uncertainty associated with the poor resolution of IRAS in the far-infrared did not allow
us to determine the distribution of far-infrared emission as well as the molecular gas in these galaxies. Without that 
information, it is difficult to clarify the nature of these unusual ULIRGs.
In this paper we present high angular resolution CO(J=1-0) observations and 
new K'-band imaging to study the widely separated ULIRGs. CO emission and the 
molecular gas are known to be associated with the source of far-infrared emission. 
Thus observations of CO emission will allow us to determine the location and the distribution 
of far-infrared luminosity of these ULIRGs, in addition to the distribution of molecular gas. 
Also making use of the optical imaging data (Kim et al. 2001) and our new K'-band
images, we discuss the nature of these widely separated ULIRGs.   
We will use $q_0=0.5$ and $H_0=75$ km s$^{-1}$ Mpc$^{-1}$ throughout this paper.
\section{Observations}
We used the BIMA (Berkeley-Illinois-Maryland Association) array (Welch et al. 1996) to search for the
CO(J=1-0) emission line from the selected ULIRGs.
Each galaxy in our sample was observed in various runs during 1998 and 1999 (Table. 1) in
the B-configuration with baselines ranging from 5 k$\lambda$ to
80 k$\lambda$, and the C-configuration with baselines ranging from 3 k$\lambda$ to 30 k$\lambda$.
The digital correlator was configured to cover a velocity range of 1600 km s$^{-1}$ centered at the
velocity of the ULIRGs calculated from their redshift.   
Phase calibration was achieved by observing nearby quasars every 20 to 30 minutes. We also observed planets to determine
the absolute flux scale.

We used the MIRIAD data reduction package (Sault et al. 1994) to calibrate and process our data. After 
phase calibration and flux calibration the visibilities were Fourier-transformed to make the images. 
These images were then cleaned using the 
standard task CLEAN of the MIRIAD package. The final maps have angular resolutions of about 4 arcsec for 
B-configuration and 8 arcsec for C-configuration, respectively.

The K'-band (2.1 $\mu m$) images of 4 galaxies IRAS F11180+1623, IRAS
F14394+5332, IRAS F17028+5817, IRAS F23327+2913 were obtained with the QUIRC 1024 x 1024 near-infrared camera
on July 15, 2000, at the UH 88" telescope on Mauna Kea.
The plate scale of the QUIRC is 0.1886$^{\prime\prime}$/pixel with f/10 and the seeing is estimated about 
0.5$^{\prime\prime}$ to 0.7$^{\prime\prime}$.
The sky frames were constructed by median filtering of 9 dithered 
(20$^{\prime\prime}$ rectilinear motion) object frames.
Dome flats were obtained by turning a lamp in the dome on and off
at the beginning and end of each night.
The dark subtracted and normalized flats were used for flat-fielding the sky-subtracted images.
Flat-fielded images were then coadded after shifting in fractional
pixel units with respect to the object center.
The final coadded images were calibrated from observations
of infrared standard stars (Elias et. al. 1982).
K'-band images of two other galaxies (IRAS F10485-1447, IRAS F12359-0725) together with R-band images of all
the galaxies in our sample were taken from the work of Kim et al. (2001).
The optical spectra of the nothern object of IRAS F23327$+$2913 were taken in October 1995 at the UH 88" telescope with
spectral resolution of 8 $\AA$ in the spectral range of $\lambda\lambda$ 4500-8500$\AA$. The spectra
were processed with the standard data reduction procedures.
\section{Results}
We detected CO(J=1-0) emission from 4 galaxies IRAS F11180$+$1623, IRAS
F14394$+$5332, IRAS F17028$+$5817, IRAS F23327$+$2913  at $\geq$ 3$\sigma$ in the channel maps. 
In IRAS F12359-0725 we found possible CO(J=1-0) emission of 3$\sigma$ with channel width of 200 km s$^{-1}$ which is
coincident within the synthesized beam with one of the galaxy in the system.   
Fig 1. shows the integrated intensity maps for the galaxies observed in our 
program together with their R-band and K'-band images.  The CO emission line profiles of the detected nuclei
are shown in Fig. 2. Some of the properties of these ULIRGs are provided in Table. 2.
The CO emitting regions are not resolved by our synthesized
beams. In the case of IRAS F14394$+$5332, which is the only ULIRG mapped in B-array, we can infer an upper limit to the size 
of the CO emitting region to be $\sim$ 1" or $\sim$ 2 kpc in linear scale.

To estimate the molecular gas content from the total flux of CO(J=1-0) line we adopt the same conversion 
factor as for the Milky Way galaxy (Bryant \& Scoville 1999):
\begin{equation}
{\rm M}_{\rm g} = 1.20\times 10^4 {\rm F}_{\rm CO} {\rm D}_{\rm L}^2 (1 + z)^{-1}
\end{equation}
where M$_{\rm g}$ is the molecular gas mass in M$_\odot$, 
F$_{\rm CO}$ is the integrated flux in Jy km s$^{-1}$, D$_{\rm L}$ is the luminosity distance in Mpc, 
and $z$ is the redshift. The molecular gas mass 
derived for the galaxies in our sample is similar to that found in other ultraluminous galaxies (Solomon et al. 1997) and in 
some high redshift galaxies (Combes et al. 1999) using the same Galactic conversion factor as ours. The 1$\sigma$ upper
limit of the CO flux is rms*$\Delta V_{\rm FWHM}$/$\sqrt{\Delta V_{\rm FWHM}/\Delta V_{\rm res}}$ where rms is the noise level in the channel maps, 
$\Delta V_{\rm FWHM}$ is the linewidth determined from the gaussian fitting to the line profiles and 
$\Delta V_{\rm res}$ is the channel width. The 1$\sigma$ upper limit to
the molecular gas mass of the companion is then derived following the above equation.
For the ULIRGs with detected CO emission, the companion may possess a molecular gas content of 
up to 10\% to 20\% the molecular gas content of the main component (Table. 2).     
The upper limit is in the range of several 10$^9$ M$_\odot$ which implies the companions are not as gas rich as the main
component but may possess a molecular gas content comparable to that of normal spiral galaxies. 

Gaussian fitting to the line
profiles gives the FWHM of 200 to 400 km s$^{-1}$ except for the case of IRAS F11180$+$1623. The weak CO(J=1-0) emission from
IRAS F11180$+$1623 did not allow us to determine accurately its linewidth. Thus, the value of 600 km s$^{-1}$ should be regarded as
upper limit to the true linewidth of IRAS F11180$+$1623.

Although we did not succeed to detect CO(J=1-0) emission from IRAS F10485$-$1447, the 1$\sigma$ upper limit to 
the molecular gas mass of $\sim$ 4$\times$10$^9$ M$_\odot$ is still quite high.  In addition, this galaxy has been
observed only with the B-array which provides long baselines data with high phase scattering due to the atmospheric fluctuation. 
It would be very useful to search again for CO emission with improved sensitivity.\\
Using optical spectra, Veilleux et al. (1999a) determined the spectral type of all ULIRGs in the 1 Jy sample. 
In all cases, the galaxy detected in CO(J=1-0) line has LINER or Seyfert II spectral type while the companion has no
emission line except for IRAS F17028$+$5817E galaxy, which has spectral type HII, typical of starburst galaxies (Table 3.).\\  
The magnitudes of each component of the ULIRGs are derived from our K'-band images and presented in Table. 3. In most cases the
galaxies have compact light distribution in K'-band (Fig. 1). 
However, K'-band image of IRAS F14394$+$5332 shows the extended emission with
distortion in the E component, suggestive of a recent merger. 
\section{Detailed Description of Individual System}
{\bf IRAS F10485$-$1447} - 
This system is close to two bright foreground stars which are visible on the upper left of both R-band and K'-band 
images (Fig. 1). On the R-band image, the E component shows short tidal tails in two opposite directions
and the W component has
a long curved tidal tail of about 10 kpc. The W component also has a distortion toward SW direction.
The R-band luminosity of the W component is only slightly brighter than that of the E component,
whereas in the K'-band the W component is more than one magnitude brighter
than the E component. The optical emission lines were detected from the W component with 
LINER spectral type (Veilleux et al. 1999a). The presence of absorption lines in the spectrum of the E galaxy
(S. Veilleux, private communication) indicates that  both galaxies are at the same redshift and are interacting.
Though we did not detect any CO emission in this system, the general trend of finding molecular gas in the
active component of the widely separated ULIRGs suggests that most of the gas could reside in 
the W component.\\
{\bf IRAS F11180$+$1623} - 
The optical morphology of the E component (Fig. 1) shows two tidal tails 
in two opposite directions whereas the W component does not show any strong disturbances. 
The spectral type of the E component is LINER (Veilleux et al. 1999a) and no detectable optical 
emission lines were observed in the W component.
The CO emission was detected only on the E component.
The E component may itself be a merger because it has two opposite tidal tails.
One of the tidal tails on the E component tends to be curved and directed to the W component.
There is also a small distortion on the east side of the W component, suggesting that the
two galaxies are interacting.
The lack of CO emission as well as the non-detection of optical emission lines in the W component 
imply that this galaxy contains only a small amount of gas and is in the early stage of tidal 
interaction with the E component. Because there is no measured redshift for the W component, 
we classify this pair as a tentative merger system.
There are 3 additional galaxies around the system, and if 
they are at the same redshift as the two galaxies considered,
then this whole system is likely an example of the Hickson compact groups (Hickson 1982).\\
{\bf IRAS F12359$-$0725} - 
On the R-band image (Fig. 1), the S component shows a curved tidal tail whereas the N component has no visible tail
except a small depression on the western part of the galaxy. The two galaxies are found to be at the same redshift and
are likely in the early phase of interaction.
A foreground star is present to the west of the system on both R-band and K'-band 
images.
The R-band magnitudes of both components are similar, 
whereas K'-band magnitude of the N component is 1.1 magnitude 
brighter than that of the S component.
The N galaxy with LINER spectral type is more active than the S component which
has HII spectral type. We found possible CO(J=1-0) emission coincident with the 
N component within the synthesized beam of 10"$\times$6". As the N component is the most active in the
system, the detection follows the same trend for other widely separated ULIRGs in our sample.\\
{\bf IRAS F14394$+$5332} - 
This system has large nuclear separation
and shows long tidal tail ($\sim$ 35 kpc, Fig. 1).
The R-band luminosity of the E component is slightly higher than that of the W component.
On K'-band, the E component is about 1.1 magnitudes brighter than the W component.
Furthermore, the E component shows strong optical emission lines and has a Seyfert II 
spectral type. In contrast, the W component shows no detectable emission lines. 
We detected CO(J=1-0) emission from the E component only.

The E component has Seyfert II spectra and bright near-infrared nucleus. Radio continuum observations at 20 cm with the VLA - 
B array (4 arcsec resolution) detected a fairly strong continuum source of 38 mJy (Crawford et al. 1996) at 
the same position of this galaxy.
This result is expected as Seyfert galaxies are known to be strong continuum sources. An obscured AGN at the center of the
E component is likely the source of most of the continuum emission.

Our new K'-band image of IRAS F14394$+$5332 (Fig. 1) under good seeing condition (0.5$^{\prime\prime}$) reveals 
that the E component has two nuclei in the center with nuclear separation of 1.29$^{\prime\prime}$.
Thus, K'-band morphology suggests that this galaxy is already a merging system.\\  
{\bf IRAS F17028$+$5817} - 
On the R-band image (Fig. 1), the W component is larger than the E component, but it has a more diffuse light distribution.
Faint emission resembling tidal tails can be seen in the W component, in the R-band images of 
Kim et al. (2001) and Murphy et al. (1996). 
On K'-band image, the W component is much brighter than the E 
component and shows a compact light distribution.
The CO emission was detected in this W component. The W component also has a very red nuclear color
as observed in the typical merger ULIRGs (Kim et al. 2001).
Strong optical emission lines detected from both components indicate that they are at the same redshift.
The spectral types of the W \& E components are LINER and HII region, respectively.\\
{\bf IRAS F23327$+$2913} -
On the R-band image (Fig. 1), the S component resembles a normal spiral with very thick bar structure,
whereas the N component looks much more disturbed with a big plume on the northern direction and 
has small protrusion on the eastern direction.
Both components have comparable R-band magnitudes.
In K'-band the S component shows a compact distribution and is brighter than the N component.
Interestingly, the optical emission lines were detected only from the S component and 
not from the much disturbed N component, except for the strong stellar absorption line of
Mg Ib $\lambda$5175{\AA} (EW (Mg Ib) = 5.5{\AA}, Fig. 3).
The spectral type of the S component is LINER and the CO emission line was detected only from this component.
Radio continuum emission at 20 cm and 6 cm has been detected in this component (Crawford et al. 1996). 
The continuum emission at 6 cm was resolved by the 4 arcsec beam of the VLA.\\
The large equivalent width of the Mg Ib absorption line which originates from the intermediate stellar population 
($\ge 10^9$ yr, Bica, Alloin, \& Schmidt 1990), and the lack of Balmer absorption lines 
(originating from stellar population of age $10^8-10^9$ yr) both suggests that
star formation activity in this galaxy has already ceased some time ago.
It is unclear whether the big northern plume on the N component is a remnant of past tidal interaction
or caused by recent tidal interaction with the S component. The latter is more likely,
since if the plume was caused by the past tidal interaction more than 10$^9$ years ago then, 
it should already have been relaxed at this time and not be observable.\\
\section{Discussion}
\subsection{Molecular Gas Mass}
The most interesting result of our observations is the detection of CO emission from only
one component in the pair of the selected ULIRGs. The molecular gas content in this 
component is quite high (several times 10$^{10}$ M$_\odot$) if we adopt the Milky Way conversion
factor. However, the value of the conversion factor has been suggested to be lower for ULIRGs (Solomon et al. 1997).
By modelling the high resolution observations of CO lines from nearby ULIRGs, Downes \& Solomon (1998) derived 
molecular gas masses which are a factor of 5 lower.

From the IRAS fluxes at 60 $\mu m$ and 100 $\mu m$ we have estimated the dust temperatures in the
ULIRGs detected with CO emission. The dust temperatures are roughly the same in these galaxies, 
$\sim$ 40 K (Table. 4). If we
assume that the gas is in thermal equilibrium with the dust, namely T$_{\rm gas}$ = T$_{\rm dust}$,
we can estimate the radius of the CO emitting region from the observed CO flux F$_{\rm CO}$ (Solomon et al. 1997):
\begin{eqnarray}
L^{'}_{\rm CO} & = & 3.25\times 10^7 \: F_{\rm CO}\nu_{\rm obs}^{-2} D^{2}_L (1 +z)^{-3} \\ 
R_{\rm CO} & = & (L^{'}_{\rm CO}/\pi T_{\rm gas} f_{\rm V} \Delta V_{\rm FWHM})^{0.5}
\end{eqnarray}
where f$_{\rm V}$ is the velocity filling factor, $\Delta$V$_{\rm FWHM}$ is the 
linewidth and $\nu_{\rm obs}$ is the observed frequency in GHz, D$_L$ is the luminosity distance in Mpc. 
For simplicity, we assume that
the velocity filling factor f$_{\rm V}$ is equal to unity. We also assume the thermalization and optical thickness 
of CO(J=1-0) line, with the
CO(J=1-0) brightness temperature equal to the gas temperature. The dynamical (virial) mass (Table. 4) 
contained inside the CO emitting region is R$_{\rm CO}\Delta V_{\rm FWHM}^2$/G, where G is the gravitational constant. 
We find that the dynamical 
masses are a factor of 3 to 4 lower than the molecular gas masses. As a result, the H$_2$-to-CO luminosity
conversion factor for ULIRGs must be lower than the Galactic value. It was suggested by Downes et al. (1993)
and Solomon et al. (1997) that in ULIRGs the dynamical mass is dominated by the molecular gas mass. In this    
case, the dynamical masses estimated above are the upper limit to the actual molecular gas mass.\\
\subsection{Far Infrared and Radio Continuum Emission}
The component containing most of the molecular gas is also the most active galaxy in the pair, having either
Seyfert II or LINER-like spectral type. The other component is less active having either a HII spectral type typical
of starburst galaxies or no detectable emission lines. Since there is a correlation  
between CO emission and the 100 $\mu m$ flux from luminous and ultraluminous galaxies (Solomon et al. 1997), 
the galaxy detected in CO is likely to be the source of
most of the far-infrared luminosity coming from the system and presumably the site of vigorous 
star formation activity. Our result is consistent with the Evans et al. (1999, 2000) who found
most of the molecular gas in several warm double-nucleus ULIRGs to be associated with only the Seyfert nuclei.
All ULIRGs in our sample are also found to be associated with 20 cm continuum sources 
(Becker et al. 1994, Condon et al. 1997). The continuum sources have flux densities ranging from
a few mJy to $\sim$ 40 mJy. For two galaxies IRAS F14394$+$5332 and IRAS F23327$+$2913, high angular continuum observations
have been reported (Becker et al. 1994, Crawford et al. 1996). The continuum sources in these two galaxies coincide with the position
of CO(J=1-0) emission. In particular, the spectral index of radio continuum emission derived from 20 cm and 6 cm
observations of IRAS F23327$+$2913 is -0.52 (Crawford et al. 1996), suggesting a non-thermal nature of the emission.
For other ULIRGs in our sample, due to the lack of observations at other frequencies, 
we could not determine the nature of the continuum
emission. However, IRAS F14394$+$5332E, a Seyfert II galaxy, appears to have stronger radio
continuum flux relative to CO luminosity, compared to other ULIRGs in this sample. Significant fraction of
the radio continuum could arise from an AGN in this galaxy.
\subsection{What is the origin of the widely separated ULIRGs ?}
From the optical and K'-band images we found evidence that the active component of IRAS F14394$+$5332
possesses two nuclei. IRAS F14394$+$5332 and 3 other systems identified by Kim et al. (2001) in
the 1 Jy sample of ULIRGs by imaging and spectroscopic data can be classified as multipled mergers.
Recent survey with the HST images of 
113 ULIRGs by Borne et al. (2000) found that the fraction of multiple merger among ULIRGs might be as high as 20\%.
However, this percentage could be overestimated since Borne et al. (2000) did not identify real interacting members
with spectroscopic observations. Studies of the ULIRGs and luminous infrared galaxies (LIG), which are less
luminous in the infrared than ULIRGs, by Gao (1996, 1997) also found that ULIRGs and LIGs have 
strong clustering. They suggested that some LIGs and ULIRGs were formed by the merger of groups of gas-rich galaxies.

Other systems studied in our paper appear to be in the early stage of interaction. Thus, these systems do not fit into the
standard interpretation of the ULIRGs because the star-formation activity is expected to happen at the late stage of interaction.
Murphy et al. (1996) also found 7 ULIRGs in their sample with separation greater than 10 kpc. They suggest several 
possible scenarios to explain these particular systems such as the presence of a third nucleus from previous encounter
or the ULIRG phenomenon can occur early in the interaction. If the second scenario applies to the ULIRGs considered here, 
our detection of molecular gas in only one component of the system
indicates that the ULIRG phenomenon might be triggered during the early phase of interaction in a system 
comprised of a gas-rich galaxy and a much less gas-rich galaxy.  

The IRAS F23327$+$2913 system is unique in the whole 1 Jy sample of the ULIRGs compiled by Kim \& Sanders (1998).
The presence of absorption features in the optical spectrum of the northern component suggests that the N component 
did not experience star formation episode for the last $10^9$ yrs at least.
The molecular gas is detected in the Southern component which has a LINER spectrum. Thus, spectroscopic 
evidence suggests that the S component is more active than the N companion. 
More surprisingly, in the R-band image the S component appears as a spiral galaxy with a large bar-like structure at the center. Therefore, the morphology of
the S component is very different from other ULIRGs which have either elliptical appearance or strongly disturbed morphology (Sanders et al. 1988,
Borne et al. 2000, Kim et al. 2001). Numerical simulations of the merger involving two massive disk galaxies (Mihos \& Hernquist 1996, Springel 2000) showed
that the two galaxies are strongly disturbed during the interacting phase and the final product of the merger resembles an elliptical galaxy.
Therefore, IRAS F23327$+$2913 does not fit into the commonly accepted scenario of merger between two disk galaxies for ULIRGs. 
Other mechanisms should be considered to explain the presence of the large reservoir of molecular gas as well as the huge far-infrared 
luminosity of the S component. The extended 6 cm radio continuum emission reported by Crawford et al. (1996) indicates that
the star formation activity is not concentrated in the nuclear region as in other ULIRGs (Downes \& Solomon 1998, Bryant \&
Scoville 1999) but spread over the galaxy.
The existence of the thick bar in a system of two merging disk galaxies can 
be observed in the early stage of the tidal interaction in the numerical simulations (eg. Noguchi 1988, Barnes 1992, Barnes \& Hernquist 1996).
If this is the case, then we are witnessing the earliest stage of interaction in ULIRGs.
In order to become an ULIRG at this early phase of the interaction, the S component
should have a large amount of gas initially or the timescale to collect gas into the nuclear region is very short.
In addition, the presence of a bar-like structure suggests that the molecular gas might be driven 
to the center of the galaxy under the influence of a non-axisymmetric gravitational potential. 
However, Ho et al. (1997) and Mulchaey \& Regan (1997) found that Seyfert and LINER galaxies do not show any significant 
preference for bars. As a result, it's still unclear 
whether a bar is effective enough to drive the large amount of molecular gas currently found in the S component.
Higher resolution observations of CO emission lines and radio continuum are needed to better understand the origin of this
galaxy. 
\section{Conclusion}
We have detected CO(J=1-0) emission line in 5 out of 6 widely separated ULIRGs selected from the complete 1 Jy sample. The molecular gas mass
present in these ULIRGs indicates that they are gas-rich. Our high angular resolution observations also reveal the concentration of
the molecular gas in only one galaxy of the pair.  
Optical spectra suggest that galaxy is also the most active in the pair, having either LINER or Seyfert II spectral type. 
Using optical and K'-band images of the
ULIRGs, we found evidence in support of the multiple merger scenario for IRAS F14394+5332. Other systems are likely
in the early phase of interaction.
In particular, optical image of IRAS F23327$+$2913 reveals that a large amount of molecular gas is 
concentrated in a spiral-like galaxy with no evidence for strong tidal interaction.
Our observation results suggest that 
the conventional view of ULIRG phenomenon as the final phase of the merger of two massive disk galaxies 
probably does not apply to the widely separated ULIRGs. Other mechanisms are needed 
to explain their origin. Future sensitive and high resolution observations will help to clarify the
nature of these unusual ULIRGs.  
\acknowledgments

We would like to express our thanks to an anonymous referee for critical 
comments which are very helpful in improving our paper. We also thank Dr. S. Veilleux
for providing us the data prior to publication.

\begin{figure*}[ht]
\setlength{\unitlength}{1cm}
\begin{picture}(16.0,20.0)
\put(0,13){\resizebox{16cm}{!}{\includegraphics*{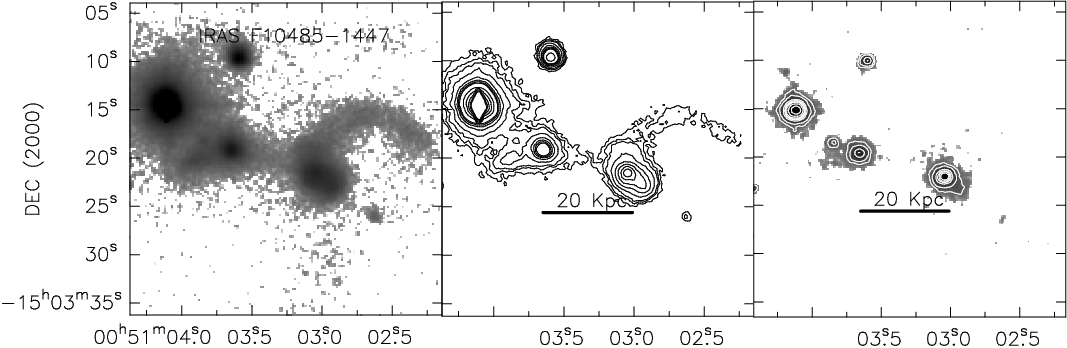}}}
\put(0,6.85){\resizebox{16cm}{!}{\includegraphics*{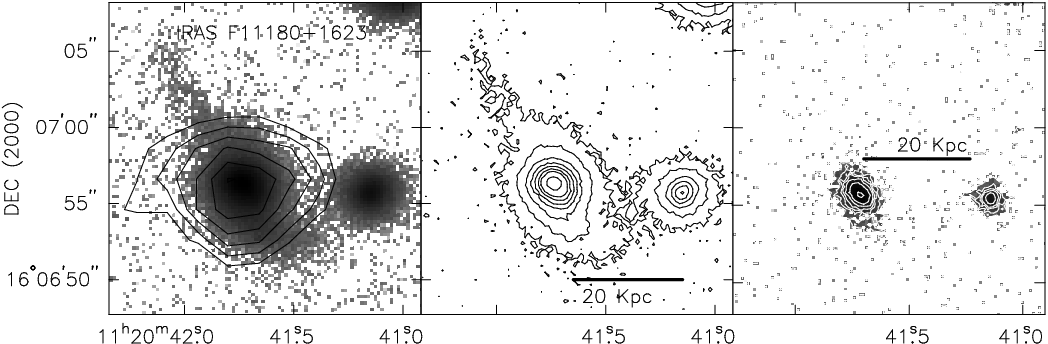}}}
\put(0,0){\resizebox{16cm}{!}{\includegraphics*{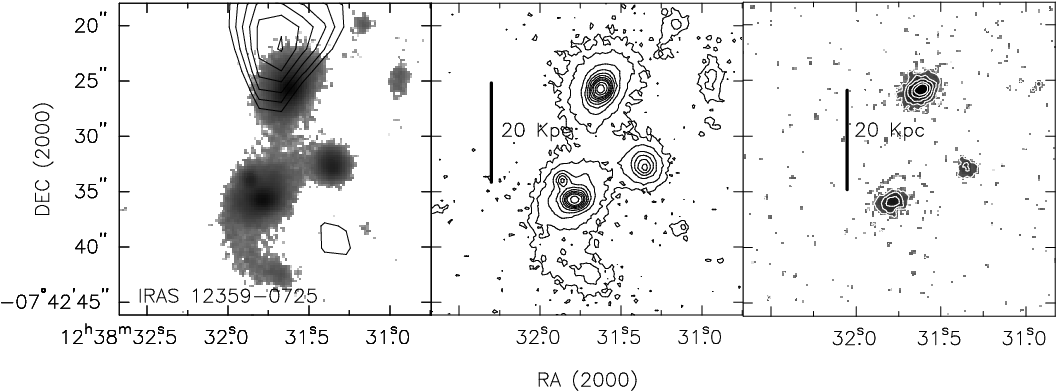}}}
\end{picture}
\caption{Left frame: the contour plots of the CO(J=1-0) emission in ULIRGs overlayed on the
R-band images. The first CO(J=1-0) contour is at 50\% of the peak. Central frame: contour plot 
of the same R-band images at the same scale. Right frame: K'-band images.}
\end{figure*}

\addtocounter{figure}{-1}

\begin{figure*}[ht]
\setlength{\unitlength}{1cm}
\begin{picture}(16.0,20.0)
\put(0,13){\resizebox{16cm}{!}{\includegraphics*{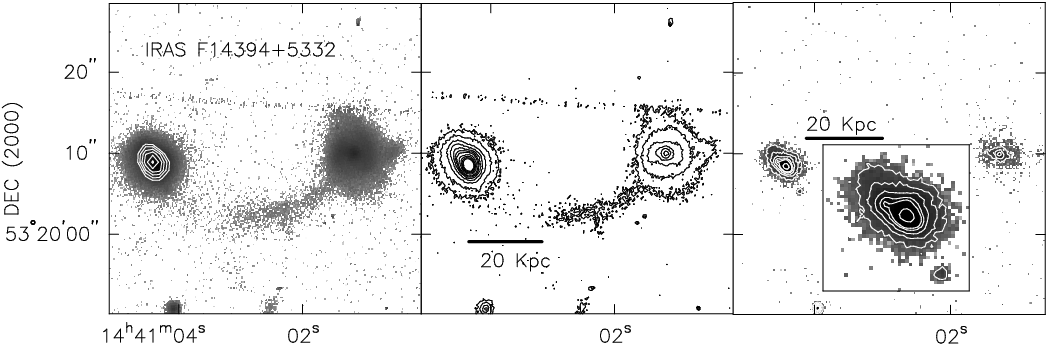}}}
\put(0,6.85){\resizebox{16cm}{!}{\includegraphics*{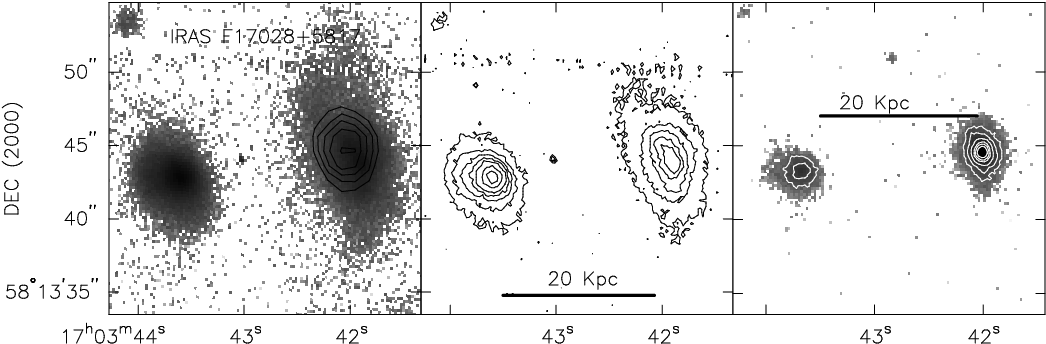}}}
\put(0,0){\resizebox{16cm}{!}{\includegraphics*{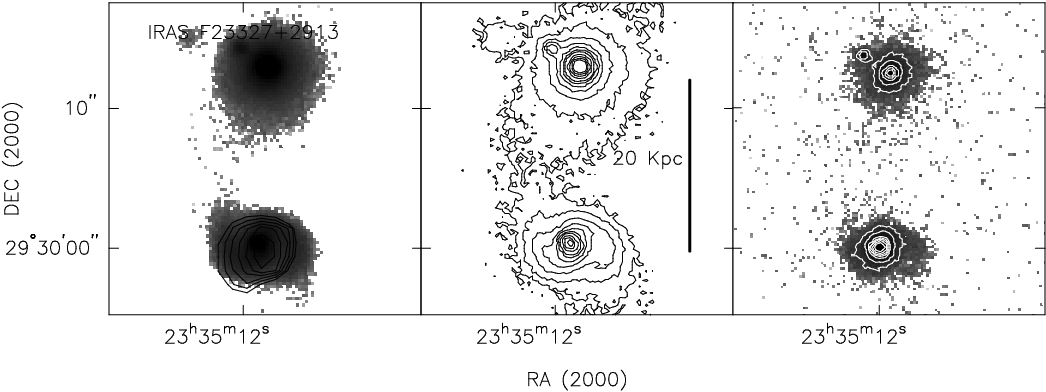}}}
\end{picture}
\caption{Continued. Left frame: the contour plots of the CO(J=1-0) emission from ULIRGs in our sample overlayed on their
R-band images. Central frame: contour plot of the same R-band images at the same scale. Right frame: K'-band images.
The small image superposed on the right frame of IRAS F14394+5332 is the K'-band blown-up image of the E component.}
\end{figure*}

\begin{figure*}[ht]
\setlength{\unitlength}{1cm}
\begin{picture}(16.0,12.0)
\put(0,5.){\resizebox{6cm}{!}{\includegraphics*{s1118.ps}}}
\put(6.5,5.){\resizebox{5.5cm}{!}{\includegraphics*{s1235.ps}}}
\put(12.5,5.){\resizebox{5.5cm}{!}{\includegraphics*{s1439.ps}}}
\put(0,0){\resizebox{6cm}{!}{\includegraphics*{s1702.ps}}}
\put(6.5,0){\resizebox{5.5cm}{!}{\includegraphics*{s2332.ps}}}
\end{picture}
\caption{The CO(1-0) line profiles from 5 ULIRGs. The velocity
is relative to the galaxy's velocity measured from optical spectra.}
\end{figure*}

\begin{figure*}[ht]
\setlength{\unitlength}{1cm}
\begin{picture}(10.0,10.0)
\put(0,-2){\resizebox{16cm}{!}{\includegraphics*{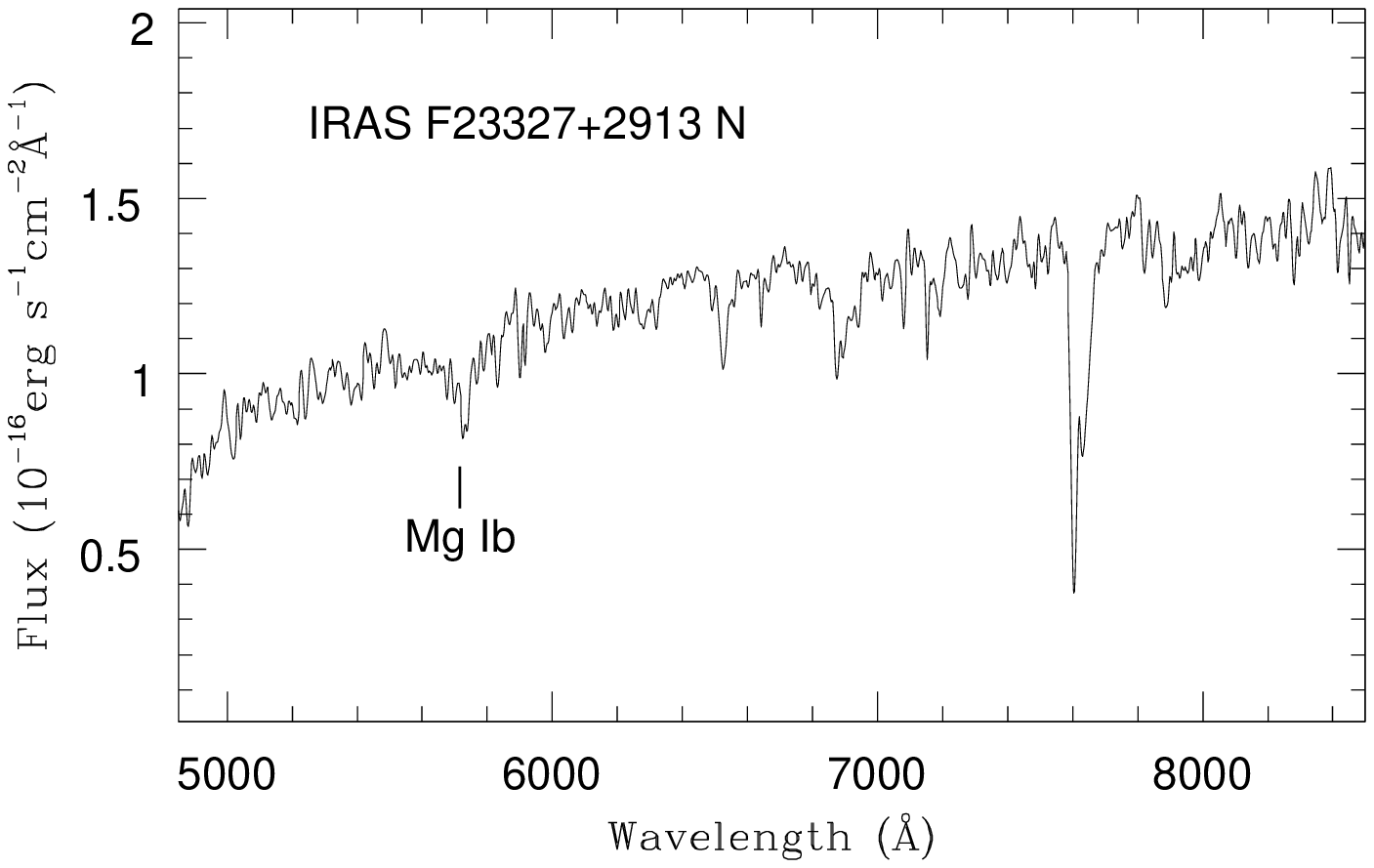}}}
\end{picture}
\caption{Optical spectra of IRAS F23327$+$2913 N taken from the University of Hawaii 2.2m 
telescope on Mauna Kea Observatory.  \label{fig3}}
\end{figure*}

\begin{table}
\begin{center}
\footnotesize
\caption{{\footnotesize Summary of the observations.}}
\vspace{0.5cm}
\begin{tabular}{lcccl}
\tableline
Name & Redshift & Obs. Freq & Num. of tracks & remarks \\
     &          &  (GHz)    &                &          \\ \hline
IRAS F10485$-$1447 & 0.133 & 101.739 & 4 B-array tracks & no detection \\ 
                   &       &         &                  &              \\ \hline
IRAS F11180+1623   & 0.166 & 98.860  & 4 B-array tracks & no detection \\
                   &       &         & 2 C-array tracks & CO(J=1-0) detection \\ \hline 
IRAS F12359$-$0725 & 0.138 & 101.291 & 3 B-array tracks & no detection \\ 
                   &       &         & 1 C-array tracks & possible detection of CO(J=1-0)\\ \hline
IRAS F14394+5332   & 0.105 & 104.317 & 4 B-array tracks & CO(J=1-0) detection \\
                   &       &         & 3 C-array tracks & CO(J=1-0) detection \\ \hline
IRAS F17028+5817   & 0.106 & 104.222 & 4 B-array tracks & no detection \\
                   &       &         & 2 C-array tracks & CO(J=1-0) detection \\ \hline
IRAS F23327+2913   & 0.106 & 104.160 & 1 B-array track  & no detection \\
                   &       &         & 3 C-array tracks & CO(J=1-0) detection \\
\tableline 
\end{tabular}
\end{center}
\end{table}

\begin{table}
\begin{center}
\footnotesize
\caption{{\footnotesize Observational properties.}}
\vspace{-0.25cm}
\begin{tabular}{cccccccc}
\tableline
Name &\multispan{2} Coordinates (J2000.0)$^a$ & D$_L$ & Beam & Total CO flux & M$_{\rm H_2}$  & $\Delta$M$_{\rm H_2}$$^b$ \\
     & R. A. & Dec. & (Mpc) & (${''}\times{''}$) & (Jy km s$^{-1}$) & (10$^{10}$ M$_\odot$) & (10$^{10}$ M$_\odot$) \\
\tableline
IRAS F10485$-$1447 & 10$^{\rm h}$51$^{\rm m}$03$^{\rm s}$.42 & -15$^{\rm o}$03'19".8 & 548.6  &  5.5$\times$4.3  &$\le$ 1  & $\le$ 0.4 & 0.4 \\
IRAS F11180+1623 & 11$^{\rm h}$20$^{\rm m}$41$^{\rm s}$.52 & 16$^{\rm o}$06'56".6 &  689.5  & 10.2$\times$9.1     &  11 & 5.4 & 0.8 \\
IRAS F12359$-$0725 & 12$^{\rm h}$38$^{\rm m}$31$^{\rm s}$.67 & -07$^{\rm o}$42'32".1 & 569.8  & 10.5$\times$6.6  & 6$^c$   & 2 & 0.4 \\
IRAS F14394+5332 & 14$^{\rm h}$41$^{\rm m}$02$^{\rm s}$.86 & 53$^{\rm o}$20'07".9 &  430.5  & 4.6$\times$3.8  & 21 & 4.2 & 0.4 \\
IRAS F17028+5817 & 17$^{\rm h}$03$^{\rm m}$42$^{\rm s}$.75 & 58$^{\rm o}$13'44".5 &  433.7  & 6.6$\times$5.5  & 14 & 2.8 & 0.3 \\
IRAS F23327+2913 & 23$^{\rm h}$35$^{\rm m}$11$^{\rm s}$.87 & 29$^{\rm o}$30'05".8 &  438.9  & 7.2$\times$5.6  & 8 & 1.7  & 0.2 \\
\tableline 
\end{tabular}
\end{center}
$^a$ From Kim et al. (2000).\\
$^b$ 1$\sigma$ upper limit of the molecular gas mass of the companion.\\
$^c$ Weak emission at 3$\sigma$ level in 200 km s$^{-1}$ channel maps.
\end{table}
\begin{table}
\begin{center}
\footnotesize
\caption{{\footnotesize Physical parameters of ULIRGs.}}
\vspace{0.5cm}
\begin{tabular}{llccccccc}
\tableline
Name  & Spectral type$^a$ & z & Proj. Separation  & F$_{\rm IR}$ & F$_{\rm 20cm}$  & \multispan{2} Total magnitude  \\
      &               &              & (kpc) & (10$^{12}$ L$_\odot$) & (mJy) & R-band  &  K'-band \\
\tableline
IRAS F10485-1447E  & absorption lines  &  0.134$^b$  &   &    &   & 18.1 & 15.9 \\
IRAS F10485-1447W  & LINER         & 0.133 & \raisebox{2.0ex}[0cm]{20.0}   & \raisebox{2.0ex}[0cm]{1.48}  & \raisebox{2.0ex}[0cm]{4.4} & 17.7 & 14.7 \\
\hline
IRAS F11180+1623E  & LINER         & 0.166 &  &   &  & 17.9 & 14.7 \\
IRAS F11180+1623W  &  no emission lines  & -  &\raisebox{2.0ex}[0cm]{21.8}  & \raisebox{2.0ex}[0cm]{1.74}  & \raisebox{2.0ex}[0cm]{4.7} & 19.4 & 16.1 \\
\hline
IRAS F12359-0725N  & LINER         & 0.138 &    &    &   & 18.2 & 14.8 \\
IRAS F12359-0725S  & HII           & 0.138 &\raisebox{2.0ex}[0cm]{22.9}    & \raisebox{2.0ex}[0cm]{1.29}   & \raisebox{2.0ex}[0cm]{4.5} & 18.2 & 15.9 \\
\hline
IRAS F14394+5332E  & Seyfert II    & 0.105 &  &   & 38.0  & 16.3 & 13.1 \\
IRAS F14394+5332W  & no emission lines  & -  &\raisebox{2.0ex}[0cm]{50.5}  & \raisebox{2.0ex}[0cm]{1.10}   &  -  & 16.9 & 14.2 \\ \hline
IRAS F17028+5817E  & HII           & 0.106 &  &   &  & 17.4 & 14.8 \\
IRAS F17028+5817W  & LINER         & 0.106 &\raisebox{2.0ex}[0cm]{23.3}  & \raisebox{2.0ex}[0cm]{1.26}   & \raisebox{2.0ex}[0cm]{16.7} & 17.1 & 13.5 \\
\hline
IRAS F23327+2913N  & absorption lines  & 0.106 &  &   & - & 16.5 & 14.1 \\
IRAS F23327+2913S  & LINER         & 0.107 &\raisebox{2.0ex}[0cm]{22.7}  & \raisebox{2.0ex}[0cm]{1.15}   & 8.4 & 16.8 & 13.6 \\
\tableline 
\end{tabular}
\end{center}
$^a$ From Veilleux, Kim, \& Sanders (1999).\\
$^b$ S. Veilleux, private communication
\end{table}
\pagebreak

\begin{table}
\begin{center}
\footnotesize
\caption{{\footnotesize Dynamical masses derived from CO data.}}
\vspace{0.5cm}
\begin{tabular}{ccccccccc}
\tableline
Name  & \multispan{2} $f_{\nu}(\lambda)^\dagger$ & T$_{\rm dust}$ & L$^{'}_{\rm CO} $       & $\Delta$V$_{\rm FWHM}^*$ & Radius & Dynamical mass \\ \cline{2-3}
      &  60 $\mu m$ & 100 $\mu m$  & (K) & (10$^9$K km s$^{-1}$pc$^2$) & (km s$^{-1}$) & (kpc)  & (10$^9$ M$_\odot$) \\
\tableline
IRAS F11180+1623 & 1.19  & 1.60 & 41.5 & 11.0 & 600 & 0.38 & 31.0 \\
IRAS F14394+5332 & 1.95  & 2.39 & 43.3 & 8.6 & 380 & 0.40 & 13.3 \\
IRAS F17028+5817 & 2.43  & 3.91 & 36.9 & 5.8 & 330 & 0.40 & 10.0 \\
IRAS F23327+2913 & 2.1   & 2.81 & 37.9 & 3.4 & 200 & 0.38 & 3.5 \\
\tableline 
\end{tabular}
\end{center}
$^\dagger$ From Kim \& Sanders (1998)\\
$^*$ The FWHM is derived from gaussian fitting of the line profile
\end{table}

\end{document}